\documentstyle[epsfig,psfig,pra,aps,eqsecnum]{revtex}
\def\be{\begin{eqnarray}}
\def\ee{\end{eqnarray}}
\def\l{\langle}
\def\r{\rangle}

\begin{document}
\author{
V. Bu\v{z}ek$^{1,2}$ and  M. Hillery$^{3}$
}
\address{
$^{1}$ Institute of Physics, Slovak Academy of Sciences,
       D\'ubravsk\'a cesta 9, 842 28 Bratislava, Slovakia\\
$^{2}$ Department of Mathematics and Physics, Comenius University,
       Mlynsk\`a dolina F2, 842 15 Bratislava, Slovakia\\
$^{3}$ Department of Physics and Astronomy, Hunter College, CUNY,
       695 Park Avenue, New York, NY 10021, USA
}
\title{UNIVERSAL OPTIMAL CLONING OF QUBITS AND QUANTUM REGISTERS
\thanks{To be presented at 
 the {\bf First NASA Conference on Quantum
Computing and Quantum Communications}, 17-20 February 1998, Palm Springs,
U.S.A.
}
}

\date{31 December 1997}
\maketitle
\begin{abstract}
We review   our recent work on the universal (i.e. input state independent)
optimal quantum copying
(cloning) of  qubits. 
We present unitary transformations which
describe the optimal cloning of a qubit 
and we present the corresponding  quantum
logical network. We also
present network  for
an optimal quantum copying ``machine'' (transformation)
which produces	$N+1$ identical copies from the original qubit.
Here again the quality (fidelity)
of the copies does not depend on the state of the original
and is only a function of the number of copies, $N$.
In addition, we present the  machine
which universaly and optimally 
clones states of quantum objects in arbitrary-dimensional Hilbert spaces.
In particular, we discuss universal cloning of quantum registers.

\end{abstract}

\section{Introduction}
The most fundamental
difference between classical and quantum information is
that while classical information can be copied perfectly,
quantum information cannot. In particular, it follows from the
{\it no-cloning theorem} \cite{Wootters}
(see also \cite{Diekes,Barnum}) that one
cannot create a perfect duplicate of an {\it arbitrary} qubit.
For example, using the well-known {\it teleportation protocol}
\cite{Bennett}, one can create a perfect copy of the
original qubit but this will be at the expense of the
{\it complete} destruction of information encoded in
the original qubit.  In contrast, the
main goal of quantum copying is to produce a copy
of the original qubit which is as close as possible to the
original state while the output state of the original qubit
is minimally disturbed.

We have shown recently \cite{Buzek1,Buzek1a}
 that if one is only interested in producing imperfect copies, 
then it is
possible to make quantum clones of the original qubit.
To be specific:
The copy machine considered by Wootters and Zurek
\cite{Wootters}
in their proof of the no-cloning theorem, for example,
produces two identical copies at its output, but
the quality of these copies depends upon the input state.
They are perfect for the basis vectors which we denote
as $|0\rangle$
and $|1\rangle$, but, because the copying process
destroys the off-diagonal information of the
input density matrix,
they are poor for input states of the form $(|1\rangle +
e^{i\varphi}|0\rangle )/\sqrt{2}$, where $\varphi$ is arbitrary.
We have introduced \cite{Buzek1,Buzek1a,Buzek3}
 a  different copying machine, 
the Universal Quantum Copying Machine
(UQCM), which produces two identical copies whose quality is
{\it independent} of the input state.  In
addition, its performance is, on average,
better than that of the
Wootters-Zurek machine, and the action of the machine
simply scales the expectation values of relevant observables.
This UQCM was shown to be optimal, in the sense that it maximizes
the average fidelity between the input and output qubits, by
Gisin and Massar \cite{Gisin} and by Bru\ss et al. \cite{Bruss}.
Gisin and Massar
have also been able to find copying transformations which produce
$N$ copies from $M$ originals (where $N>M$) \cite{Gisin}.  In 
addition, we have proposed 
quantum logic newtworks for quantum copying machines 
 \cite{Buzek3,Buzek4}, and bounds have been placed on
how good copies can be \cite{Hillery,Ekert}.

In this talk we will firstly  review our original ideas on
universal quantum copying of a single qubit (Section II). In Section III
we will present a quantum network describing the UQCM.
Secondly, in Section IV we will  introduce the copying machine
which produces $N+1$ identical copies
from the original qubit.   The quality (fidelity)
of copies does not depend on the state of the original
and is only a function of a number $N$ of produced  copies.
We present a  quantum network for the quantum copying machine.
We show that this machine is formally described by the
same unitary transformation as recently introduced  by
Gisin and Massar \cite{Gisin}. In Section V we will analyze
properties of multiply cloned qubits.
Thirdly, in Section VI we show how quantum registers
(i.e. systems composed of many entangled qubits) can  be
universally cloned. To be specific, one approach is to use the
single-qubit copiers to copy individually (locally) each qubit.
We have shown earlier \cite{Buzek2} that in the case of two qubits this
local copying will preserve some of quantum correlations between qubits,
but as we will show, it does not make a particularly good copy of the
two-qubit state. As an alternative we propose a copy machine which
universally (and optimally) clones quantum states in arbitrary-dimensional
Hilbert spaces. This allow us to discuss optimal cloning of quantum
registers.

\section{Universal quantum copying machine}
Let us assume we want to copy
an arbitrary pure state $|\Psi\rangle_{a_0}$ which in a
particular basis
$\{|0\rangle_{a_0},|1\rangle_{a_0}\}$ is described by
the state vector $|\Psi\rangle_{a_0}$
\begin{eqnarray}
|\Psi\rangle_{a_0} = \alpha |0\rangle_{a_0}
+\beta |1\rangle_{a_0};
\qquad \alpha=\sin\vartheta/2 {\rm e}^{i\varphi};
~~~\beta=\cos\vartheta/2.
\label{1.1}
\end{eqnarray}
The two numbers which characterize the state (\ref{1.1}) can be
associated with the ``amplitude'' $|\alpha|$
and the ``phase'' $\varphi$ of the qubit. Even though
 ideal copying, i.e., the transformation
$|\Psi\rangle_{a_0} \longrightarrow |\Psi\rangle_{a_0}
|\Psi\rangle_{a_1}$
is prohibited by the laws  of quantum mechanics for an
{\em arbitrary} state (\ref{1.1}), it is still possible
to design quantum copiers which
operate reasonably well. In particular, the
UQCM \cite{Buzek1} is specified by the following conditions.

{\bf (i)}The  state of the original system and
its quantum copy at the
output of the quantum copier, described by density operators
$\hat{\rho}^{(out)}_{a_0}$ and $\hat{\rho}^{(out)}_{a_1}$,
respectively, are identical, i.e.,
\begin{eqnarray}
\hat{\rho}^{(out)}_{a_0}   = \hat{\rho}^{(out)}_{a_1}
\label{1.3}
\end{eqnarray}

{\bf (ii)} If no {\em a priori} information about the
{\em in}-state of the original system is available,
then it is reasonable to require
that {\em all} pure states should be copied equally
well. One way to implement
this assumption is to design a quantum
copier such that the distances between
density operators of each system at the output
($\hat{\rho}^{(out)}_{a_j}$
where $j=0,1$)	and the ideal density operator
$\hat{\rho}^{(id)}$
which describes the {\em in}-state of the original
mode are input state
independent.  Quantitatively this means that if we employ
the Bures distance
\cite{Bures}
\begin{eqnarray}
d_B(\hat{\rho}_1,\hat{\rho}_2)
=\sqrt{2}\left(1 -{\rm Tr}\sqrt{\hat{\rho}_1^{1/2}\hat{\rho}_2
\hat{\rho}_1^{1/2}}\right)^{1/2},
\label{1.4}
\end{eqnarray}
as a measure of distance between two operators,
then the quantum copier should be such that
\begin{eqnarray}
d_{B}(\hat{\rho}_{a_j}^{(out)};\hat{\rho}_{a_j}^{(id)})=
{\rm const.};\qquad j=0,1.
\label{1.5}
\end{eqnarray}

{\bf (iii)}
Finally, we would also like to require that
the copies are as close as possible to the ideal
output state, which is, of course, just the input state.
This means that we want our quantum copying transformation
to minimize the distance between the output state $\hat{\rho}_{a_j}^{(out)}$
of the copied qubit and the ideal  state $\hat{\rho}_{a_j}^{(id)}$. The
distance is minimized with respect to all possible unitary transformations
$U$ acting on the Hilbert space ${\cal H}$ of two qubits and
the quantum copying machine
(i.e., ${\cal H}={\cal H}_{a_0}\otimes{\cal H}_{a_1}\otimes{\cal H}_{x}$)
\begin{eqnarray}
d_B(\hat{\rho}_{a_j}^{(out)};\hat{\rho}_{a_j}^{(id)}) =
{\rm min}
\left\{d_B^{(U)}(\hat{\rho}_{a_j}^{(out)};
\hat{\rho}_{a_j}^{(id)}); \forall U\right\};
 \qquad (j=0,1).
\label{1.6}
\end{eqnarray}
Originally, the UQCM was found by analyzing a transformation
which contained two free parameters, and then determining
them by demanding that condition (ii) be satisfied, and that the
distance between the two-qubit output density matrix and the
ideal two-qubit output be input state independent.  That the
UQCM machine obeys the condition (\ref{1.6}) has only been shown
recently \cite{Gisin,Bruss}.

The unitary transformation which implements the
UQCM \cite{Buzek1} is given by
\begin{eqnarray}
|0\rangle_{a_0} |Q\rangle_{x} &\rightarrow &
 \sqrt{\frac{2}{3}}|00
\rangle_{a_0a_1}|\uparrow\rangle_{x}+\sqrt{\frac{1}{3}}|
+\rangle_{a_0a_1}
|\downarrow\rangle_{x} \nonumber \\
|1\rangle_{a_0} |Q\rangle_{x} &\rightarrow &
\sqrt{\frac{2}{3}}|11
\rangle_{a_0a_1}|\downarrow\rangle_{x}
+\sqrt{\frac{1}{3}}|+\rangle_{a_0a_1}
|\uparrow\rangle_{x},
\label{1.7}
\end{eqnarray}
where
$|+\rangle_{a_0a_1} = 
(|10\rangle_{a_0a_1}+|01\rangle_{a_0a_1})/\sqrt{2}$,
and satisfies the conditions (\ref{1.3}-\ref{1.6}).
The system labelled by $a_0$ is the original (input) qubit,
while the other system $a_1$ represents the qubit onto which
the information is copied. This qubit is supposed to be prepared
initially in a state $|0\rangle_{a_1}$ (the ``blank paper''
in a copier).  The states of the copy machine
are labelled by $x$.  The state space of the copy
machine is two dimensional, and we assume that it is always
in the same state $|Q\rangle_{x}$ initially.  If the
original qubit is in the superposition state  (\ref{1.1})
then the reduced density operator of both copies
at the output are equal [see condition (\ref{1.3})]
and they can be expressed as
\begin{equation}
\hat{\rho}_{a_j}^{(out)}
=\frac{5}{6}|\Psi\rangle_{a_j}\langle\Psi|+
\frac{1}{6}|\Psi_{\perp}\rangle_{a_j}
\langle\Psi_{\perp}|,\qquad j=0,1
\label{1.9}
\end{equation}
where
$|\Psi_{\perp}\rangle_{a_j}=\beta^{\star}
|0\rangle_{a_j}-
\alpha^{\star} |1\rangle_{a_j}$ ,
is the state orthogonal to $|\Psi\rangle_{a_j}$.  This
implies that the copy contains $5/6$ of the state we
want and $1/6$ of the one we do not.

The density operator $\rho_{a_j}^{(out)}$ given by
Eq.(\ref{1.9})
can be rewritten in a ``scaled'' form:
\begin{equation}
\hat{\rho}_{a_j}^{(out)}
= s_j \hat{\rho}_{a_j}^{(id)}
+ \frac{1-s_j}{2} \hat{1};\qquad j=0,1,
\label{1.11}
\end{equation}
which guarantees that the distance (\ref{1.4})
is input-state independent,
i.e. the condition (\ref{1.5} is
automatically fulfilled. The scaling
factor in Eq.(\ref{1.11}) is $s_j=2/3$ ($j=0,1$).

We note that once again that the UQCM copies
all input states with the
same quality and therefore is suitable for copying when
no {\it a priori} information about the state of the
original qubit is available. This corresponds to a uniform
prior probability distribution
on the state space of a qubit (Poincare sphere).
Correspondingly, one can measure the quality of copies
by the fidelity ${\cal F}$, which is equal to the mean
overlap between a copy and the input state
\cite{Gisin}
\begin{equation}
{\cal F}=\int\, d\Omega  _{a_j}\langle \Psi|
\hat{\rho}_{a_j}^{(out)}| \Psi \rangle_{a_j},
\label{1.12}
\end{equation}
where $\int d\Omega=\int_0^{2\pi} d\varphi\,
\int_0^{\pi} d\vartheta
\sin\vartheta/4\pi$. It is easy to show that the relation
between the fidelity ${\cal F}$
and the scaling factor is
$s= 2{\cal F}-1$.

\section{Copying network}

In what follows we show how, with simple quantum
logic gates, we can copy
quantum information encoded in the original
qubit onto other qubits.
The copying procedure can be understood as a ``spread''
of information
via a ``controlled'' entanglement between
the original qubit and the
copy qubits. This controlled entanglement is
implemented by a sequence
of controlled-NOT operations operating on the
original qubit and the
copy qubits which are initially prepared in a
specific state.

In designing a network for the UQCM we first
note that since
the state space of the copy machine itself is
two dimensional, we can
consider it to be an additional qubit.	Our network,
then, will take
3 input qubits (one for the input, one which
becomes one the copy,
and one for the machine) and transform them
into 3 output qubits.
In what follows we will denote
the quantum copier qubit as $b_1$ rather than $x$.
The operation of this network is such, that in order
to transfer information from the original $a_0$
qubit to the target qubit $a_1$
we will need one  {\em idle} qubit $b_1$ which
plays the role of quantum copier.

Before proceeding with the network itself let
us specify the one and
two-qubit gates from which it will be constructed.
Firstly we define a single-qubit rotation
$\hat{R}_j(\theta)$
which acts on the basis vectors of qubits as
\begin{eqnarray}
\hat{R}_{j}(\theta) | 0 \rangle_j   = 
\cos\theta | 0 \rangle_j + \sin\theta | 1 \rangle_j;\qquad
\hat{R}_{j}(\theta) | 1 \rangle_j   = 
-\sin\theta | 0 \rangle_j + \cos\theta | 1 \rangle_j.
\label{2.1}
\end{eqnarray}

We also will utilize a two-qubit operator
(a two-bit quantum gate), the so-called
controlled-NOT gate, which has as its inputs a
control qubit (denoted as $\bullet$ in Fig.1) and
a target qubit	(denoted as $\circ$ in Fig.1).
The control qubit
is unaffected by the action of the
gate, and if the control qubit is
$|0\rangle$, the target qubit is
unaffected as well.  However, if the
control qubit is in the
$|1\rangle$ state, then a NOT operation is
performed on the target
qubit.	The operator which implements this
gate, $\hat{P}_{kl}$,
acts on the basis vectors of the two
qubits as follows ($k$
denotes the control qubit and $l$ the target):
\begin{eqnarray}
\begin{array}{c}
\hat{P}_{kl} | 0 \rangle_k | 0 \rangle_l  = 
| 0 \rangle_k | 0 \rangle_l;\qquad
\hat{P}_{kl} | 0 \rangle_k | 1 \rangle_l  = 
| 0 \rangle_k | 1 \rangle_l;
\\
\hat{P}_{kl} | 1 \rangle_k | 0 \rangle_l  = 
| 1 \rangle_k | 1 \rangle_l;\qquad
\hat{P}_{kl} | 1 \rangle_k | 1 \rangle_l  = 
| 1 \rangle_k | 0 \rangle_l.
\end{array}
\label{2.2}
\end{eqnarray}

We can decompose the quantum copier network into two parts.
In the first part the copy ($a_1$) and the idle ($b_1$)
qubits	 are prepared in a specific state
$|\Psi\rangle_{a_1b_1}^{(prep)}$. Then in the second part
of the copying network the original information from the
original qubit $a_0$
is {\em redistributed} among the three qubits.	  That
is the action of the quantum copier can be described as
a sequence of two unitary transformations
\begin{eqnarray}
|\Psi\rangle_{a_0}^{(in)} |0\rangle_{a_1}
|0\rangle_{b_1}
\longrightarrow  |\Psi\rangle_{a_0}^{(in)}
|\Psi\rangle_{a_1 b_1}^{(prep)}
\longrightarrow  |\Psi\rangle_{a_0a_1b_1}^{(out)}.
\label{2.3}
\end{eqnarray}
The network for the quantum copying machine is displayed
in Fig.\ 1.

\subsection{Preparation of quantum copier}
Let us first look at the preparation stage.
Prior to any interaction with the input qubit we have to
prepare the two quantum copier qubits ($a_1$ and $b_1$)
in a  specific state $|\Psi\rangle_{a_1b_1}^{(prep)}$
\begin{eqnarray}
|\Psi\rangle_{a_1b_1}^{(prep)} =
\frac{1}{\sqrt{6}}\left(
2 |00\rangle_{a_1b_1} + |01\rangle_{a_1b_1}
+ |11\rangle_{a_1b_1}\right),
\label{2.8}
\end{eqnarray}
which
can be prepared by a simple quantum network
(see the ``preparation''
box in Fig.1) with two controlled-NOTs $\hat{P}_{kl}$
and three  rotations $\hat{R}(\theta_j)$, i.e.
\begin{eqnarray}
|\Psi\rangle_{a_1b_1}^{(prep)}=
\hat{R}_{a_1}(\theta_3) \hat{P}_{b_1a_1}
\hat{R}_{b_1}(\theta_2) \hat{P}_{a_1b_1}
\hat{R}_{a_1}(\theta_1)|0\rangle_{a_1}|0\rangle_{b_1},
\label{2.6}
\end{eqnarray}
with the rotation angles defined as \cite{Buzek3}
\begin{eqnarray}
\cos 2\theta_1 = \frac{1}{\sqrt{5}};\qquad
\cos 2\theta_2 = \frac{\sqrt{5}}{3};\qquad
\cos 2\theta_3= \frac{2}{\sqrt{5}}.
\label{2.9}
\end{eqnarray}
\begin{figure}
\begin{center}
\epsfig{width=15.0truecm,file=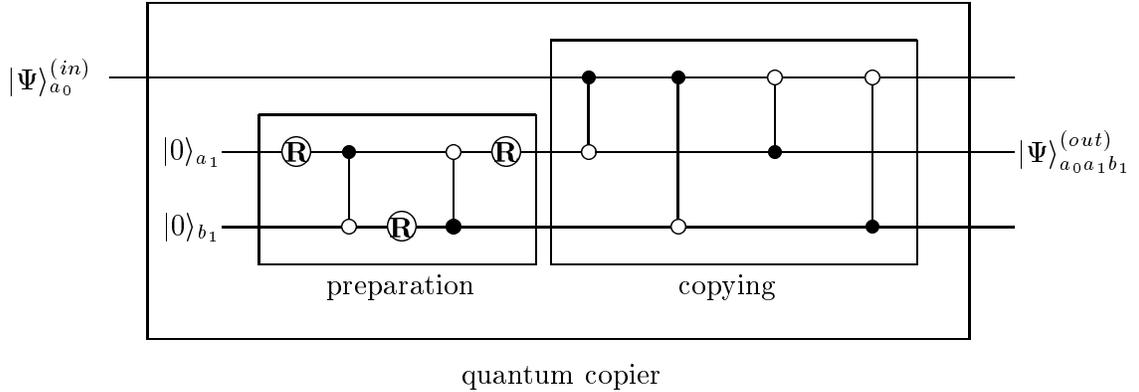}
\end{center}
\caption{
Graphical representation of the UQCM network.  The logical
controlled-NOT $\hat{P}_{kl}$ given by Eq.(\ref{2.2})
has as its input a control qubit
(denoted as $\bullet$ ) and  a target qubit  (denoted as $\circ$ ).
The action of the single-qubit operator {\bf R} is specified by the
transformation (\ref{2.1}).
We separate the preparation of the quantum copier from the
copying process itself. The copying, i.e. the transfer of quantum
information from the original qubit, is performed by a sequence of
four controlled-NOTs. We note that the amplitude information from the
original qubit is copied in the obvious direction in an XOR or
the controlled-NOT
operation. Simultaneously, the phase information is copied in the
opposite direction making the XOR a simple model of quantum non-demolition
measurement and its back-action.
}
\end{figure}

\subsection{Quantum copying}

Once the qubits of the quantum copier are
properly prepared then the copying of the initial  state
$|\Psi\rangle_{a_0}^{(in)}$
of the original qubit can be performed
by a sequence of four controlled-NOT operations (see
Fig. 1)
\begin{eqnarray}
|\Psi\rangle_{a_0a_1b_1}^{(out)} =
\hat{P}_{b_1a_0} \hat{P}_{a_1a_0}
\hat{P}_{a_0b_1}
\hat{P}_{a_0a_1}|\Psi\rangle_{a_0}^{(in)}
|\Psi\rangle_{a_1b_1}^{(prep)}.
\label{2.10}
\end{eqnarray}
When this operation is combined with
the preparation stage, we find
that the basis states of the original
qubit ($a_0$) are copied as
described by Eq.(\ref{1.7}) with
$| \uparrow\rangle_x\equiv |0\rangle_{b_1}$
and $| \downarrow\rangle_x\equiv |1\rangle_{b_1}$.
When the original qubit is in the superposition state
(\ref{1.1}) then the state vector of the three qubits
after the copying has been performed reads
\begin{eqnarray}
|\Psi\rangle_{a_0a_1b_1}^{(out)}=
|\Phi_0\rangle_{a_0a_1}|0\rangle_{b_1}
+ |\Phi_1\rangle_{a_0a_1}|1\rangle_{b_1},
\label{2.13}
\end{eqnarray}
with
\begin{eqnarray}
|\Phi_0\rangle_{a_0a_1}=
\alpha\sqrt{\frac{2}{3}}|00\rangle_{a_0a_1}
+\beta \frac{1}{\sqrt{3}}|+\rangle_{a_0a_1}; ~~
|\Phi_1\rangle_{a_0a_1}
=\beta\sqrt{\frac{2}{3}}|11\rangle_{a_0a_1}
+\alpha \frac{1}{\sqrt{3}}|+\rangle_{a_0a_1}.
\label{2.14}
\end{eqnarray}
From this it follows that at
the output of the  quantum copier we
find a pair of entangled
qubits in a state described by the density operator
\begin{eqnarray}
\hat{\rho}_{a_0a_1}^{(out)}=
|\Phi_0\rangle_{a_0a_1}\langle \Phi_0|+
|\Phi_1\rangle_{a_0a_1}\langle \Phi_1|.
\label{2.15}
\end{eqnarray}
Each of the copy qubits at the output of the quantum copier
has a reduced density operator $\hat{\rho}_{a_j}^{(out)}$
($j=0,1$)  given by  Eq.(\ref{1.11}). The Bures distance
$d_B(\hat{\rho}_{a_j}^{(out)};\hat{\rho}_{a_j}^{(id)})$ ($j=0,1$)
between the output qubit and the ideal qubit is constant and it reads
\begin{eqnarray}
d_B(\hat{\rho}_{a_j}^{(out)};\hat{\rho}_{a_j}^{(id)})
=\sqrt{2}\left(1-\sqrt{\frac{5}{6}}\right).
\label{2.16}
\end{eqnarray}

We note, that the idle qubit after the copying is performed is in
a state
\begin{eqnarray}
\hat{\rho}_{b_1}^{(out)}=
\frac{1}{3}\left(\hat{\rho}_{b_1}^{(id)}\right)^{\rm T}
+ \frac{1}{3}\hat{1},
\label{2.18}
\end{eqnarray}
where the superscript T denotes the transpose.

Finally we note that the flow of quantum information in our
network can be effectively controled by the choice of the 
preparation of the quantum copier. That is, we can a priori decide
which qubits at the output will be clones of the original
qubit \cite{Buzek1a,Buzek3}.

\section{Multiple copying}
Here we present a generalization of the transformation
(\ref{1.7}) to the case when a set of
$N$ copy qubits $a_j$ ($j=1,..,N$) are produced out of the
original qubit $a_0$. We also present 
a simple quantum network
which realizes this multiple quantum copying
$1\rightarrow 1 +N$ \cite{Buzek4}.

We already know that ideal multiple copying of the form
$|\Psi\rangle_{a_0} \longrightarrow |\Psi\rangle_{a_0}
|\Psi\rangle_{a_1}
... |\Psi\rangle_{a_N}$
does not exist. But, as we shall show, one can
generalize the copying
procedure described in Section III, and find a
transformation such that
\begin{eqnarray}
\hat{\rho}^{(out)}_{a_0}
= \hat{\rho}^{(out)}_{a_j},\qquad j=1,...,N,
\label{3.2}
\end{eqnarray}
with the distances $d_B$ [see Eq.(\ref{1.4})] which do not
depend on the initial state (\ref{1.1}) of the original qubit.

To find the $1\longrightarrow 1+N$ network we assume
the following:\newline
{\bf (1)} We assume that the information
from the original qubit
is copied to $N$ copy qubits $a_j$ which
are initially prepared in the state
$|N;0\rangle_{\vec{a}}\equiv |0\rangle_{a_1}...
|0\rangle_{a_N}$
(here the subscript $\vec{a}$ is a shorthand
notation indicating that
$|N;0\rangle_{\vec{a}}$ is a vector in the
Hilbert space of $N$ qubits $a_j$).\newline
{\bf (2)} To implement multiple quantum copying we need
to associate an  {\em idle} qubit $b_j$ with
each copy qubit, $a_j$.
These $N$ idle qubits, which play the role of
the copying machine itself,
are initially prepared in the state
$|N;0\rangle_{\vec{b}}\equiv |0\rangle_{b_1}...
|0\rangle_{b_N}$.
\newline
{\bf (3)} Prior  to the transfer  of information
from the original
qubit, the copy and the idle qubits
have been prepared
in a specific state
$|\Psi\rangle^{(prep)}_{\vec{a}\vec{b}}$. Once
this is done the copying is performed by a simple
sequence of controlled-NOT operations.

\subsection{Preparation of the quantum copier}

In order to find the explicit form for
the quantum network for $1\rightarrow
1+N$ copying we introduce normalized  state vectors
$|N;k\rangle_{\vec{a}}$ describing a
{\em symmetric} $N$-qubit state
with $k$ qubits in the state
$|1\rangle$ and $(N-k)$ qubits in the state
$|0\rangle$. 
These states are orthonormalized, i.e.
$_{\vec{a}}\langle N;l|N;k\rangle_{\vec{a}} =\delta_{k,l}$, 
and have the property
\begin{eqnarray}
|N;l\rangle_{\vec{a}}=
\sqrt{\frac{N-l}{N}}|0\rangle_{a_m}
|N-1;l\rangle_{a_1...a_{m-1}a_{m+1}...a_{N}}
+\sqrt{\frac{l}{N}}|1\rangle_{a_m}|N-1;l-1
\rangle_{a_1...a_{m-1}a_{m+1}...a_{N}}.
\label{3.5}
\end{eqnarray}
As we have already said, we assume that
the copy+idle qubits are initially prepared
in the state
\begin{eqnarray}
|\Psi\rangle^{(in)}_{\vec{a}\vec{b}}=
|N;0\rangle_{\vec{a}} |N;0\rangle_{\vec{b}}.
\label{3.6}
\end{eqnarray}
By performing a sequence
of local rotations {\bf R} and controlled-NOT
operations analogous to Eq.(\ref{2.6}) we can obtain the state
$|\Psi\rangle^{(prep)}_{\vec{a}\vec{b}}$ \cite{Barenco}
\begin{eqnarray}
|\Psi\rangle^{(prep)}_{\vec{a}\vec{b}}=
\sum_{k=0}^N \left[\,
e_k |N;k\rangle_{\vec{a}} + f_k |N;k-1\rangle_{\vec{a}}\right]
|N;k\rangle_{\vec{b}},
\label{3.7}
\end{eqnarray}
where
\begin{eqnarray}
e_k=\sqrt{\frac{2}{N+2}}\frac{
\left(\begin{array}{c} N\\k \end{array}\right)}
{\left(\begin{array}{c} N+1\\k \end{array}\right)}
;\qquad
f_k=\sqrt{\frac{k}{N-k+1}} e_k.
\label{3.8}
\end{eqnarray}
Once the copying machine is prepared in the state
$|\Psi\rangle^{(prep)}_{\vec{a}\vec{b}}$ we can start to copy
information from the original qubit $a_0$.

\subsection{Copying of information}
To describe the copying network we firstly introduce an operator
$\hat{Q}_{a_0\vec{a}}$ which is a product of the controlled-NOTs
defined by Eq.(\ref{2.2}) with $a_0$ being a control qubit and
$a_j$ ($j=1,...,N$) being targets:
\begin{eqnarray}
\hat{Q}_{a_0\vec{a}}\equiv
\hat{P}_{a_0a_N} \hat{P}_{a_0a_{N-1}} ... \hat{P}_{a_0a_1}.
\label{3.9}
\end{eqnarray}
We also introduce the operator $\hat{Q}_{\vec{a}a_0}$ describing
the controlled-NOT process with $a_0$ playing the role of the
target qubit, i.e.
\begin{eqnarray}
\hat{Q}_{\vec{a}a_0}\equiv
\hat{P}_{a_Na_0} \hat{P}_{a_{N-1}a_0} ... \hat{P}_{a_1a_0}.
\label{3.10}
\end{eqnarray}
\begin{figure}
\begin{center}
\epsfig{width=7.0truecm,angle=-90,file=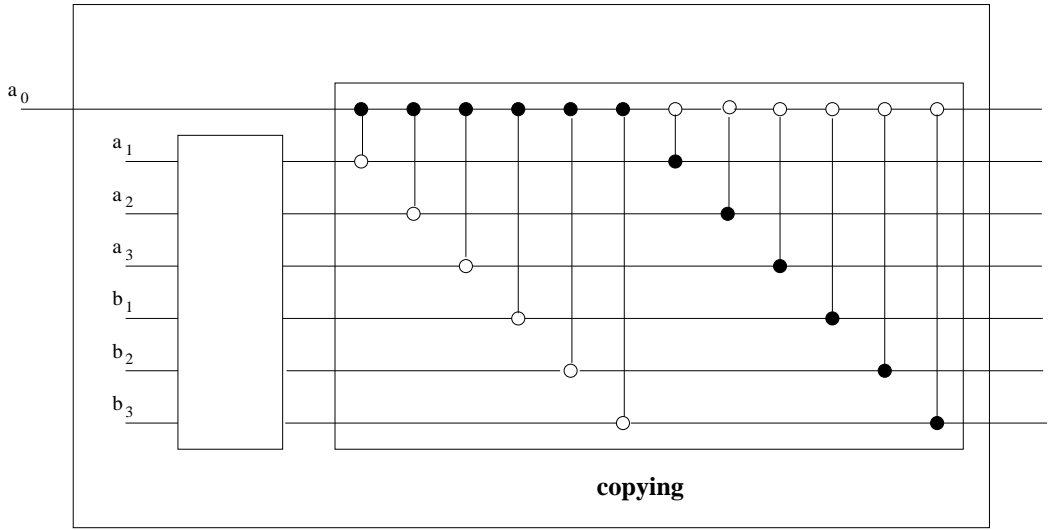}
\end{center}
\caption{
Graphical representation of the network for the $1\rightarrow 1+N$
copying.
The logical
controlled-NOT $\hat{P}_{kl}$ given by Eq.(\ref{2.2})
has as its input a control qubit
(denoted as $\bullet$ ) and  a target qubit  (denoted as $\circ$ ).
We separate the preparation of the quantum copier from the
copying process itself. The copying, i.e. the transfer of quantum
information from the original qubit, is performed by a sequence of
controlled-NOTs as described by Eq.(\ref{3.12}).
}
\end{figure}

Now we	find the $1\longrightarrow 1+N$ copying network to be
\begin{eqnarray}
|\Psi\rangle_{a_0}^{(in)} |N;0\rangle_{\vec{a}}
|N;0\rangle_{\vec{b}}
\longrightarrow  |\Psi\rangle_{a_0}^{(in)}
|\Psi\rangle_{\vec{a}\vec{b}}^{(prep)}
\longrightarrow  |\Psi\rangle_{a_0\vec{a}\vec{b}}^{(out)},
\label{3.11}
\end{eqnarray}
where the $(2N+1)$ qubit output of the copying
process is described
by the state vector
$|\Psi\rangle_{a_0\vec{a}\vec{b}}^{(out)}$
which is defined as
\begin{eqnarray}
|\Psi\rangle_{a_0\vec{a}\vec{b}}^{(out)}=
\hat{Q}_{\vec{b}a_0} \hat{Q}_{\vec{a}a_0}
\hat{Q}_{a_0\vec{b}}
\hat{Q}_{a_0\vec{a}}
|\Psi\rangle_{a_0}^{(in)}.
|\Psi\rangle_{\vec{a}\vec{b}}^{(prep)}.
\label{3.12}
\end{eqnarray}
This last equation describes a simple quantum network when
firstly the original qubit controls the target qubits of the
quantum copier. Then  the qubits
$\vec{a}$ and $\vec{b}$ ``control'' the state of the
original qubit
via another sequence of controlled-NOTs (see Fig.2).
In this way one can produce out
of a single original
qubit a set of quantum clones.

\section{ Properties of copied qubits}

Using the explicit expression for the  output state
$|\Psi\rangle_{a_0\vec{a}\vec{b}}^{(out)}$
we find that the
original and the copy qubits at the output of
the quantum copier
are in the same state described by the density operator
\begin{equation}
\hat{\rho}_{a_j}^{(out)}
= s^{(N)} \hat{\rho}_{a_j}^{(id)}  + \frac{1-s^{(N)}}{2}
\hat{1};\qquad j=0,1,...,N,
\label{4.1}
\end{equation}
where the scaling factor $s^{(N)}$ depends
on the number $N$ of
copies, i.e.
\begin{equation}
s^{(N)}=\frac{1}{3}+\frac{2}{3(N+1)},
\label{4.2}
\end{equation}
which corresponds to the fidelity ${\cal F}=2/3+1/3(N+1)$.
We see that this result for $N=1$ reduces to the case of the
UQCM discussed in Section III. We also note that in the limit
$N\rightarrow\infty$, i.e. when an infinite number of copies
is {\em simultaneously} produced
via the generalization of the UQCM,
the copy qubits still carry information about the
original qubit, because their density operators are given by
the relation
\begin{equation}
\hat{\rho}_{a_j}^{(out)}
= \frac{1}{3} \hat{\rho}_{a_j}^{(id)}  + \frac{1}{3}
\hat{1};\qquad j=0,1,...,\infty,
\label{4.3}
\end{equation}
which corresponds to the fidelity ${\cal F}=2/3$.
This is the optimal
fidelity achievable when an {\it optimal} measurement
is performed on a single
qubit \cite{Massar,Derka}. From this point of
view one can consider
quantum copying as a transformation of
quantum information into
classical information \cite{Gisin}.
This also suggests  that
quantum copying can be utilized to obtain
novel insights into the quantum
theory of measurement [e.g.,  a simultaneous
measurement of conjugated
observables on two copies of the original
qubit; or a specific
 realization of  the generalized  (POVM) measurement  performed
on the original qubit.

{\bf Comment 1}\newline
We note that if the original qubit is copied
sequentially by a
system of $N$ copying machines of the type
$1\rightarrow 1+1$
(each machine copies two outcomes of
the previous copier) then
$2^N$ copies of the original qubit in the
limit $N\rightarrow\infty$
are in the state $\hat{\rho}^{(out)}_{a_j}=
\hat{1}/2$.
In this case the copied qubits do not carry
information about the
original qubit, while all idle qubits are
in the state (\ref{2.18}).

{\bf Comment 2}\newline
Using the copying transformation (\ref{3.12})
we find that the
basis vectors $|0\rangle_{a_0}$ and $|1\rangle_{a_0}$
of the original qubit are transformed as
[compare with Eq.(\ref{1.7})]
\begin{eqnarray}
|0\rangle_{a_0} |\Psi\rangle_{\vec{a}\vec{b}}^{(in)}
&\rightarrow & \sum_{k=0}^N \lambda_k^{(N+1)}
|N+1;k\rangle_{a_0\vec{a}} |N;k\rangle_{\vec{b}};
\nonumber \\
|1\rangle_{a_0} |\Psi\rangle_{\vec{a}\vec{b}}^{(in)}
&\rightarrow & \sum_{k=0}^N \lambda_{N-k}^{(N+1)}
|N+1;k+1\rangle_{a_0\vec{a}} |N;k\rangle_{\vec{b}},
\label{4.6}
\end{eqnarray}
where
$\lambda_k^{(N+1)}=[2(N+1-k)/(N+1)(N+2)]^{1/2}$.
We clearly see that the set of $N+1$
completely symmetric orthonormal
states	$|N;k\rangle_{\vec{b}}$ (with $k=0,1,...,N$)
of the idle qubits $b_j$
plays the role of a set of basis vectors
of the abstract quantum copier and in
this form the transformation (\ref{4.6})
describes the action of the quantum copier
as discussed by Gisin and Massar
\cite{Gisin}. These authors have also
shown that transformation (\ref{4.6}) describes
the {\em optimal} input-state independent
$1\rightarrow 1+N$ quantum copier.

{\bf Comment 3}\newline
We note that idle qubits $b_j$ after the
copying is performed are always in the state
\begin{eqnarray}
\hat{\rho}_{b_j}^{(out)}=
\frac{1}{3}\left(\hat{\rho}_{b_j}^{(id)}\right)^{\rm T}
+ \frac{1}{3}\hat{1},
\qquad j=1,...,N,
\label{4.8}
\end{eqnarray}
{\em irrespective} of
the number of copies created from the original qubit.

\subsection{Inseparability of cloned qubits}

We first recall that
a density operator of two subsystems is
inseparable if it {\em cannot} be written as a convex sum
\begin{eqnarray}
	\hat{\rho}_{xy}  = \sum_m w^{(m)} 
	\hat{\rho}_{x}^{(m)}
	\otimes \hat{\rho}_{y}^{(m)}.
	\label{18}
\end{eqnarray}
Inseparability is one of the most fundamental quantum phenomena.
It is required for a violation of Bell's inequality (to be
specific, a separable system always satisfy Bell's inequality, but the
contrary is not necessarily true). Distant parties
cannot prepare an inseparable state from a separable one if they only
use local operations and classical communication.
In the case of two spins-1/2
we can utilize the Peres-Horodecki theorem
\cite{Peres,Horodecki} which states that the positivity of the
partial transposition of a state is {\em necessary} 
and {\em sufficient} for
its separability.

{\bf Comment 4}\newline
The two-qubit density operator
$\hat{\rho}^{(out)}_{a_ma_n}$ (here
$m,n=0,1,...,N$ and $m\neq n$) associated
with the output state
$|\Psi\rangle_{a_0\vec{a}\vec{b}}^{(out)}$
[see Eq.(\ref{3.12})]
in the basis $|11\rangle_{a_ma_n},
|10\rangle_{a_ma_n}, |01\rangle_{a_ma_n},
|00\rangle_{a_ma_n}$
is described by the matrix
\begin{eqnarray}
\hat{\rho}_{a_ma_n}^{(out)} =\frac{1}{6}
\left(
\begin{array}{cccc}
\frac{(3N+5) |\beta|^2 + (N-1) |\alpha|^2}{N+1} &
\frac{\alpha^{\star}\beta (N+3)}{N+1} &
\frac{\alpha^{\star}\beta (N+3)}{N+1} & 0\\
\frac{\alpha\beta^{\star} (N+3)}{N+1} & 1 & 1 &
\frac{\alpha^{\star}\beta (N+3)}{N+1} \\
\frac{\alpha\beta^{\star} (N+3)}{N+1} & 1 & 1 &
\frac{\alpha^{\star}\beta (N+3)}{N+1} \\
0 & \frac{\alpha\beta^{\star} (N+3)}{N+1} &
\frac{\alpha\beta^{\star} (N+3)}{N+1} &
\frac{(3N+5) |\alpha|^2 + (N-1) |\beta|^2}{N+1}
\end{array}
\right).
\label{4.4}
\end{eqnarray}
From Eq.(\ref{4.4}) we find that the
eigenvalues $\vec{E}=\{E_1,E_2,E_3,E_4\}$ of
the partially transposed matrix
$(\hat{\rho}_{a_ma_n}^{(out)})^{T_2}$  are
input-state independent  and  read:
\begin{equation}
\vec{E}=\left\{\frac{1}{6},\frac{1}{6},
\frac{1}{3}+\frac{\sqrt{2(5+4N+N^2)}}{6(N+1)},
\frac{1}{3}-\frac{\sqrt{2(5+4N+N^2)}}{6(N+1)}\right\}.
\label{4.5}
\end{equation}
Using the Peres-Horodecki theorem \cite{Peres,Horodecki}
 we can conclude that the two copied
qubits at the output of the copier are
inseparable
only in the case $N=1$. In this case on of
the eigenvalues (\ref{4.5})
is negative, which is the necessary and
sufficient condition for
the inseparability of the matrix (\ref{4.4}).
For $N>1$ all pairs of
copied qubits  at the output of the quantum
copier	are separable  (i.e., the eigenvalues
given by Eq.(\ref{4.5}) are positive).

{\bf Comment 5}\newline
To quantify how the ``quantum copier''
(i.e., the idle qubits)
is entangled with the original and the copy
qubits at the output,
we evaluate the parameter
$\xi^{(N)}={\rm Tr}[\hat{\rho}_{\vec{b}}^{(out)}]^2$,
which quantifies the purity of the quantum copier.
If $\xi=1$, then the
copier (i.e., the subsytem of the whole system
$a_0\vec{a}\vec{b}$)
is in a pure state. Otherwise (i.e.,
when  $\xi<1$) it is in
an impure state. If the whole system is
in a pure state, i.e.
${\rm Tr}[\hat{\rho}_{a_0\vec{a}\vec{b}}^{(out)}]^2=1$,
then $\xi$
quantifies the degree of entanglement between
the two subsystems.
From Eq.(\ref{3.12}) we find
\begin{equation}
\xi^{(N)}=\frac{1}{N+1} \frac{2 (2 N^2 + 7N +6)}{3 (N+2)^2},
\label{4.11}
\end{equation}
from which it follows that in the limit of large $N$
\begin{equation}
\xi^{(N)}\simeq \frac{4}{3(N+1)}.
\label{4.12}
\end{equation}
The lower bound $\xi_{\rm min}$ of  the purity parameter
$\xi$ of an arbitrary quantum system
in the $N+1$ dimensional
Hilbert space (i.e., this is the size of the Hilbert space of the quantum
copier) is
\begin{equation}
\xi_{\rm min}= \frac{1}{N+1}.
\label{4.13}
\end{equation}
We see that  for all values of $N$ the parameter
$\xi^{(N)}$ is
very close to its lower bound, i.e. the quantum
copier and the
copies are highly entangled.
To understand the nature of this entanglement,
we briefly consider
the $1\rightarrow 1+1$ quantum copying.
In this case, we can evaluate
the density operator $\hat{\rho}^{(out)}_{a_1b_1}$
which in  matrix
form can be written as:
\begin{eqnarray}
\hat{\rho}_{a_1b_1}^{(out)} =\frac{1}{6}
\left(
\begin{array}{cccc}
4 |\beta|^2 + |\alpha|^2 &
\alpha\beta^{\star} &  2 \alpha^{\star}\beta&2 \\
\alpha^{\star}\beta  & |\beta|^2 & 0
& 2 \alpha^{\star}\beta\\
2 \alpha\beta^{\star}  & 0 & |\alpha|^2
& \alpha\beta^{\star}\\
2 & 2 \alpha\beta^{\star}  & \alpha^{\star}\beta
& 4 |\alpha|^2 + |\beta|^2
\end{array}
\right).
\label{4.14}
\end{eqnarray}
For $\alpha$ and $\beta$ real,
the eigenvalues of the corresponding
partially transposed matrix do not depend
on these parameters and
they read:
\begin{equation}
\vec{E}=\left\{\frac{1}{3},\frac{2}{3},
\frac{1-\sqrt{17}}{12},
\frac{1+\sqrt{17}}{12}\right\}.
\label{4.15}
\end{equation}
We see that one of the eigenvalues is negative which means that
each copy qubit (i.e., either $a_0$ or $a_1$) and the idle qubit
are {\em quantum-mechanically} entangled. In the case when
$\alpha$ and $\beta$ are complex, the eigenvalues of the partially
transposed matrix associated with the matrix Eq.(\ref{4.14}) do
depend on $\alpha$ and $\beta$ and one of the eigenvalues
is {\em always} negative. So these qubits are quantum-mechanically
entangled.

\section{CLONING OF QUANTUM REGISTERS}

In what follows we will propose
a copy machine which copies  universally  
higher dimensional systems.   
We shall be particularly interested in how the quality of the
copies scales with the dimensionality, $M$, of the system 
being copied.  What we find is that the fidelity of the copies
decreases with $M$, as expected, but, somewhat surprisingly, 
does not go to zero as $M$ goes to infinity.

Let us now consider a quantum system prepared in a pure state
which is described by the vector 
\be
| \Phi\r_{a_0} = \sum_{i=1}^{M}\alpha_i |\Psi_i\r_{a_0}
\label{1}
\ee
in an $M$-dimensional Hilbert space spanned by $M$ orthonormal
basis vectors $|\Psi_i\r_{a_0}$ ($i=1,...,N$). The complex
amplitudes $\alpha_i$ are normalized to unity, i.e.
$\sum |\alpha_i|^2 =1$. In particular, one can consider $M=2^m$ where
$m$ is the number of qubits in a given quantum register.
One can generalize the no-cloning theorem which has been proven for
spin-1/2 particles (qubits) by Wootters and Zurek \cite{Wootters}
for arbitrary quantum systems. That is,
there does not exist a unitary transformation such that the state 
given in Eq. (\ref{1})
can be ideally cloned (copied), i.e. it is impossible to find a
unitary transformation such that
$|\Phi\r_{a_0} \longrightarrow |\Phi\r_{a_0}|\Phi\r_{a_1}$. 

Following our previous discussion we can ask whether a
universal cloning transformation exists which will generate two 
imperfect copies from the original state, $|\Phi\r_{a_0}$.  
The quality of the cloning should not depend
on the particular state (in the given Hilbert space) which is going
to be copied. This input-state independence (invariance)
of the cloning can be formally expressed as 
\be
\hat{\rho}_{a_j}^{(out)} = s \hat{\rho}_{a_j}^{(id)} + \frac{1-s}{M}\hat{1},
\label{3}
\ee
where $\hat{\rho}_{a_j}^{(id)}=|\Phi\r_{a_{0}} \,_{a_{0}}\l\Phi|$ 
is the density operator describing the original state 
which is going to be copied.
This scaling form of Eq.(\ref{3}) guarantees that the Bures distance
(\ref{1.4}) 
between the input and the output density operators  is input-state
independent. 

The quantum copying machine we shall use is itself an
$M$ dimensional quantum system, and we shall let 
$| X_i\r_x$ ($i=1,...,M$) be an orthonormal basis of the
copying machine Hilbert space. 
This copier is initially prepared in
a particular state $|X\r_x$. The action of the cloning 
transformation
can be specified by a unitary transformation acting on 
the basis vectors of the tensor product space 
of the original quantum system $|\Psi_i\r_{a_0}$, 
the copier, and an additional $M$-dimensional system
which is to become the copy (which is initially
prepared in an arbitrary state $| 0\r_{a_1}$).  We make
the {\em Ansatz}
\be
|\Psi_i\r_{a_0}| 0 \r_{a_1} | X \r_{x}
\longrightarrow 
c |\Psi_i\r_{a_0}| \Psi_i \r_{a_1} | X_i \r_{x}
+ d\sum_{j\neq i}^M \left(
|\Psi_i\r_{a_0}| \Psi_j \r_{a_1}+
|\Psi_j\r_{a_0}| \Psi_i \r_{a_1}\right) | X_j \r_{x};
\qquad i=1,...,M,
\label{4}
\ee
and we assume that the coefficients $c$ and $d$ are real.
From the unitarity of the transformation in Eq. (\ref{4}) 
it follows that $c$ and $d$ satisfy the relatation
\be
c^2 + 2(M-1) d^2 = 1.
\label{5}
\ee
Using this transformation we find that the particles 
$a_0$ and $a_1$ at the output of the cloner are in 
the same state (have the same reduced density matrixes), 
which is described by the density operator
\be
\hat{\rho}_{a_k}^{(out)} =   \sum_{i=1}^M |\alpha_i|^2
\left(c^2 +(M-2) d^2\right) |\Psi_i\r\l\Psi_i| 
+ \sum_{\stackrel{i,j=1}{i\neq j}}^M\alpha_i\alpha_j^*
\left( 2cd + (M-2)d^2\right)  |\Psi_i\r\l\Psi_j|
+d^2 \hat{1}.
\label{6}
\ee
Now our task is to find the values for $c$ and $d$
such that the density operator in Eq. (\ref{6}) takes the 
scaled form of Eq. (\ref{3}). This directly guarantees the universality
of the transformation (\ref{4}), i.e. the fidelity of the cloning
does not depend on the initial states of the particle which is going to
be cloned.

Comparing these two equations 
we find that $c$ and $d$ must satisfy the equation
\be
c^2 = 2cd.
\label{7}
\ee
Taking into account the normalization condition in
Eq. (\ref{5}) we find that
\be
c^2 = \frac{2}{(M+1)};\qquad d^2 =\frac{1}{2(M+1)};
\label{8}
\ee
from which it follows that the scaling factor $s$ is
\be
s= c^2 + (M-2) d^2 = \frac{(M+2)}{2(M+1)}.
\label{9}
\ee
If $M=2$, then the transformation in 
Eq. (\ref{4}) reduces
to the copying transformation for qubits given
by Eq.(\ref{1.7}). From earlier results of Gisin and Massar \cite{Gisin}
then optimality of the tranformation (\ref{4}) for $M=2$ directly
follows. At the moment we are not able to prove rigorously that
the cloning tranformation (\ref{4}) is {\it optimal} for arbitrary $M>2$.
Nevertheless, we have performed numerical tests which suggest that the 
the cloning transformation (\ref{4}) is optimal.

We first note that the scaling factor, which describes the 
quality of the copy, is a decreasing function of $M$.  This
is not surprising, because a quantum state in a large dimensional  
space contains more quantum information than one in a small
dimensional one (e.\ g.\ a state in a 4 dimensional
space contains information about 2 qubits while a state
in a 2 dimensional one describes only a single qubit), 
so that as $M$ increases one is trying to copy more and
more quantum information.  On the other hand, it is interesting 
to note that in the limit $M\rightarrow\infty$, i.e.
in the case when the Hilbert space of the given quantum system
is infinite dimensional (e.g. quantum-mechanical harmonic oscillator),
the cloning can still be performed efficiently with the 
scaling factor equal to $1/2$.

In order to confirm that the quality of the copies which the 
copying transformation in Eq. (\ref{4}) produces
is input-state independent (i.e.
all states are cloned equally well) we evaluate the Bures distance
(\ref{1.4}).
In our particular case we find, that the distance between
$\hat{\rho}^{(out)}_{a_k}$ and $\hat{\rho}^{(id)}_{a_k}$ 
depends
only on the dimension of the Hilbert space $M$, but not on the
state which is cloned, i.e.
\be
d_B(\hat{\rho}^{(out)}_{a_k},\hat{\rho}^{(id)}_{a_k})
=\sqrt{2}\left(1 -\sqrt{\frac{M+3}{2(M+1)}}\right)^{1/2}.
\label{11}
\ee
The Bures distance in Eq. (\ref{11}) is  maximal  when
states in the infinite-dimensional Hilbert space are cloned, 
and in that case we find
\be
\lim_{M\rightarrow\infty} d_B(\hat{\rho}^{(out)}_{a_k},
\hat{\rho}^{(id)}_{a_k})
=\sqrt{2-\sqrt{2}}.
\label{12}
\ee
This means that even for an infinite-dimensional system,
reasonable cloning can be performed,
which is reflected in the fact that the corresponding scaling
factor $s$ is equal to $1/2$.

Now we evaluate 
the von Neumann entropy 
$S=-{\rm Tr} \hat{\rho}\ln\hat{\rho}$ of the ouput state (\ref{6}):
\be
S=\ln[2(M+1)] - \frac{M+3}{2(M+1)}\ln (M+3).
\label{13}
\ee 
This is an increasing function of $M$ which
implies that the copy states are becoming more
and more mixed as $M$ increases.
In the limit $M\rightarrow\infty$ we find that
\be
S\simeq \ln\sqrt{M}.
\label{14}
\ee
This can be compared to the maximum value of the von Neumann 
entropy in an $M$-dimensional Hilbert space which is equal to
$S=\ln M$.

Using the transformation in Eq. (\ref{4}) we can also find
the state of the copy machine after the cloning has been 
performed
\be
\hat{\rho}_{x}^{(out)} = 2d^2 \left(\hat{\rho}_{x}^{(id)}\right)^T 
+ 2 d^2\hat{1},
\label{15}
\ee
i.\ e.\  the copier is left in a state proportional
to the transposed state of the original quantum system.
The von Neumann entropy of the copier at the output reflects 
the degree of entanglement between the copies and the copier. 
As expected,
this entropy does not depend on the state to be copied and  
is just a function of the dimension of the Hilbert space, i.e.
\be
S=\ln(M+1) - \frac{2\ln 2}{M+1}.
\label{16}
\ee
This is again and increasing function of $M$ which reflects the
fact that the copies and the copier become increasingly correlated
as $M$ increases.  On the other hand,
it follows from the Araki-Lieb theorem that the maximum
value of the entropy of two entangled subsystems with the 
dimension of
the smaller subsystem being $M$, is equal to $S_{max}= \ln M^2$,
which shows that the cloner and the clones are far from 
being maximally entangled.  

\subsection{Local vs.nonlocal cloning}

Finally, we would like to compare two methods of
copying of quantum registers. In particular, we consider
cloning of 
an entangled state of two qubits.
We assume that the two qubits are prepared in the state
\be
|\Phi\r_{a_0b_0} = \alpha|00\r_{a_0b_0} +\beta|11\r_{a_0b_0},
\label{17}
\ee
where, for simplicity, we have taken $\alpha$ and $\beta$ 
to be real, and $\alpha^2+
\beta^2=1$. First, we shall consider the case in which
each of the two qubits $a_0$ and $b_0$
is copied {\em locally} by 
two independent  quantum copiers \cite{Buzek2}. 
Each of these two copiers is
described by the transformation in Eq. (\ref{4}) with $M=2$.
Next, we shall consider a {\em nonlocal} cloning of the 
two-qubit state in Eq. (\ref{17}) when this system 
is cloned via the unitary transformation in Eq. 
(\ref{4}) with $M=4$, i.e. the cloner in this case 
can act non-locally on the two qubits. 
Our chief task will be analyze how inseparability is cloned 
in these two scenarios, but we shall also examine the quality
of the copies which are produced in the two cases.
From the Peres-Horodecki 
theorem it follows that the state in Eq. (\ref{17})
is inseparable for all values
of $\alpha^2$ such that $0<\alpha^2<1$.

Now suppose that each of the two original qubits 
$a_0$ and $b_0$ is cloned by two independent
local cloners $X_I$ and $X_{II}$, each  
described by the transformation in Eq. (\ref{4}) with $M=2$.
The cloner $X_I$ ($X_{II}$) generates out of qubit
$a_0$ ($b_0$) two qubits $a_0$ and $a_1$ ($b_0$ and $b_1$). 
After we perform trace over the two cloners we obtain a 
four-qubit density operator $\hat{\rho}^{(out)}_{a_0a_1b_0b_1}$ 
which also describes
two nonlocal two-qubit systems, i.e. $\hat{\rho}_{a_0b_1}$
and $\hat{\rho}_{a_1b_0}$. These two two-qubit systems are
the clones of the original two-qubit register (\ref{17}) and they
are described by the density operators \cite{Buzek2}
\be
	\hat{\rho}^{(out)}_{a_{0}b_{1}} =
	\hat{\rho}^{(out)}_{a_{1}b_{0}} =
	\frac{24\alpha^2+1}{36} |00\rangle\langle 00|
	+\frac{24\beta^2+1}{36} |11\rangle\langle 11|
+ \frac{5}{36}(|01\rangle\langle 01|+|10\rangle\langle 10|)
	+ \frac{4\alpha\beta}{9}(|00\rangle\langle 11|+|11\rangle\langle 00|).
\label{21}
\ee
We first note that this density matrix cannot be expressed
in the scaled form of Eq. (\ref{3}), and that the quality
fo the copies depends on the input state.  Therefore, this
procedure does not produce a universal quantum copy machine.
From the Peres-Horodecki theorem we immediately find that the
density operators in Eq. (\ref{21}) are inseparable  if 
\be
\frac{1}{2} -\frac{\sqrt{39}}{16}\leq \alpha^2 \leq 
\frac{1}{2} +\frac{\sqrt{39}}{16}.
\label{22}
\ee
This proves that for a restricted set of pure two-qubit  
states (\ref{17}), those
which satisfy the condition in Eq. (\ref{22}), it is possible
to {\em locally} copy them
so that their original inseparability is (partially) preserved.

Let us now see what happens when we copy the entire two-qubit
register at once.  We would like to determine
whether the set  of original two-qubit
states (see Eq. (\ref{17})), which after the cloning exhibit 
inseparability,
is larger (i.e., the restriction of the form given
in Eq. (\ref{22}) is weaker)
than when a nonlocal cloning is performed. To do so, we introduce
four basis vectors $|\Psi_1\r=|00\r$;
$|\Psi_2\r=|01\r$;
$|\Psi_3\r=|10\r$; and  $|\Psi_4\r=|11\r$, so that the original 
two-qubit state in Eq. (\ref{17}) 
is expressed as $|\Phi\r= \alpha|\Psi_1\r + 
\beta |\Psi_4\r$. The copying is now performed according to the
transformation in Eq. (\ref{4}) with $M=4$. 
We find that each of the two pairs of two-qubit copies 
at the output of the copier
is described by the same density operator 
\be
	\hat{\rho}^{(out)}_{a_{0}b_{1}} =
	\frac{6\alpha^2+1}{10} |00\rangle\langle 00|
	+\frac{6\beta^2+1}{10} |11\rangle\langle 11|
+ \frac{1}{10}(|01\rangle\langle 01|+|10\rangle\langle 10|)
	+ \frac{3\alpha\beta}{5}(|00\rangle\langle 11|+|11\rangle\langle 00|).
\label{23}
\ee
Here the fidelity of copying is input-state independent. Moreover, the
quality of the cloned register is higher  than that in Eq. (\ref{21}).
Again, using the Peres-Horodecki theorem we find that the density
operator in Eq. (\ref{23}) is inseparable if
\be
\frac{1}{2} -\frac{\sqrt{2}}{3}\leq \alpha^2 \leq 
\frac{1}{2} +\frac{\sqrt{2}}{3}.
\label{24}
\ee
We conclude that  quantum inseparability
can be copied better (i.e. for much larger range
of the parameter $\alpha$) by using a nonlocal copier
than when two local copiers are used.

\section{Conclusions}
We have presented  the universal optimal quantum
copying machine which optimally clones  
a single original qubit to $N+1$ qubits.  We have  found a simple
quantum  network which realizes this quantum copier.
In addition we have presented a universal cloner for quantum registers.
We have numerically tested the optimality of this cloner, but the
rigorous proof has still be presented.

Quantum copiers can be
effectively utilized in various processes
designed for manipulation
with quantum information. In particular, quantum
copiers can be used for
 an optimal eavesdropping  \cite{Gisin2}; they can  be
applied for realization of the optimal
generalized (POVM) measurements
\cite{Derka2}, or they can be utilized for storage
and retrieval of
information in quantum computers \cite{DiVincenzo}.

\end{document}